\documentclass[a4paper]{article}
\usepackage{ASVspoof}
\usepackage{epsfig,amssymb,amsmath}
\usepackage{hyperref}
\usepackage{multirow}
\usepackage{algorithmic}
\usepackage{algorithm}
\usepackage[table,xcdraw]{xcolor}
\ninept

\title{SZU-AFS Antispoofing System for the ASVspoof 5 Challenge}

\name{Yuxiong Xu$^1$, Jiafeng Zhong$^1$, Sengui Zheng$^2$, Zefeng Liu$^1$, Bin Li$^1$$^*$\thanks{*Corresponding author}}

\address{
	$^1$Guangdong Provincial Key Laboratory of Intelligent Information Processing, Shenzhen Key \\
	  Laboratory of Media Security, and SZU-AFS Joint Innovation Center for AI Technology,\\ 
	  Shenzhen University, Shenzhen, China\\
	 $^2$Afirstsoft Technology Group co., Ltd., Shenzhen, China\\
{\small \tt \{xuyuxiong2022, jiafengzhong2022\}@email.szu.edu.cn, 1225620446@qq.com}\\
{\small	\tt liuzefeng2021@email.szu.edu.cn, libin@szu.edu.cn }}
\begin{document}
\maketitle

\begin{abstract}
    This paper presents the SZU-AFS anti-spoofing system, designed for Track 1 of the ASVspoof 5 Challenge under open conditions. The system is built with four stages: selecting a baseline model, exploring effective data augmentation (DA) methods for fine-tuning, applying a co-enhancement strategy based on gradient norm aware minimization (GAM) for secondary fine-tuning, and fusing logits scores from the two best-performing fine-tuned models. The system utilizes the Wav2Vec2 front-end feature extractor and the AASIST back-end classifier as the baseline model. During model fine-tuning, three distinct DA policies have been investigated: single-DA, random-DA, and cascade-DA. Moreover, the employed GAM-based co-enhancement strategy, designed to fine-tune the augmented model at both data and optimizer levels,  helps the Adam optimizer find flatter minima, thereby boosting model generalization. Overall, the final fusion system achieves a minDCF of 0.115 and an EER of 4.04\% on the \textit{evaluation} set.
\end{abstract}

\section{Introduction}
Recent advancements in Artificial Intelligence Generated Content (AIGC) have significantly enhanced the naturalness, fidelity, and variety of speech.
Unfortunately, this progress has resulted in a proliferation of forgeries that can be almost indistinguishable from authentic speech to the human auditory system.
Concurrently, automatic speaker verification (ASV) systems have become increasingly susceptible to spoofing and deepfake attacks, in which attackers produce convincingly realistic simulations of the target speaker's voice \cite{Wu2015}. 
The potential misuse of spoofed speech presents significant societal risks.
Therefore, developing a robust and generalizable anti-spoofing system to counter these threats has emerged as a critical research imperative.

The ASVspoof challenges \cite{Wu2015a,Kinnunen2017,Todisco2019,Yamagishi2021,wang2024asvspoof5} have significantly boosted interest in developing robust detection solutions for spoofing and deepfake attacks, thereby enhancing the security and reliability of ASV systems.
These challenges provide standardized benchmark protocols and comprehensive evaluation datasets.
What's more, the last four ASVspoof challenges \cite{Wu2015a,Kinnunen2017,Todisco2019,Yamagishi2021} have prompted the proposal of numerous innovative spoofing detection methods \cite{Chen2020,Monteiro2020,Wang2021,Zhang2021}.

\begin{figure}[t]
	\centering
	\includegraphics[scale=0.26]{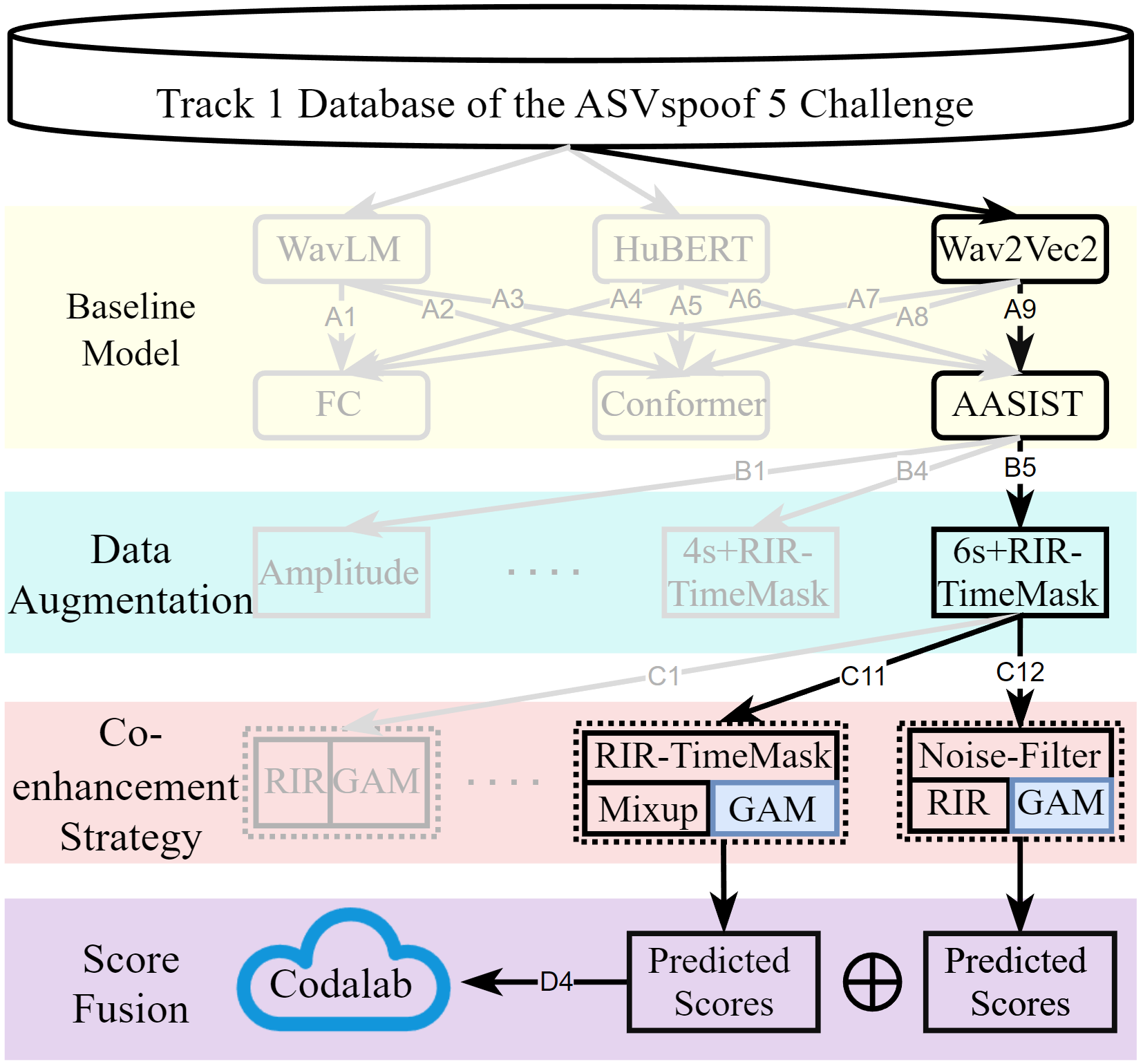} 
	\caption{Illustration of the SZU-AFS anti-spoofing system. The colored boxes represent four stages of the system, with each stage labeled by model IDs from A to D. The best-performing model in each stage and its ID number are presented in bold. First, a baseline model (A9) was selected, combining the Wav2Vec2 feature extractor with the AASIST classifier. The A9 model was then fine-tuned using the RIR-TimeMask method to obtain the best-augmented model (B5), which was subsequently further fine-tuned using a GAM-based co-enhancement strategy. Finally, the logits scores from the C11 and C12 models were fused using an average score-level fusion method, and the results were submitted for evaluation on the Codalab platform.}
	\vspace{-5pt} 
	\label{Fig1}
\end{figure}

Held in 2024, the ASVspoof 5 Challenge \cite{wang2024asvspoof5} presents two distinct conditions, open and closed, for both Track 1 which focuses on standalone speech deepfake detection, and Track 2 which is dedicated to spoofing-robust automatic speaker verification.
Under the closed conditions, participants are restricted from using specified data protocols.
Conversely, the open conditions offer greater flexibility, allowing participants to utilize external data and pre-trained self-supervised models, provided there is no overlap between training data (i.e., that used for training foundational models) and evaluation data.
Track 1 is similar to the DF track of the ASVspoof 2021 Challenge, reflecting a scenario in which an attacker has access to a targeted victim's voice data, such as data posted on social media.
In this scenario, the \textit{evaluation} set contained data processed with conventional codecs or modern neural codecs.
Track 2, similar to the LA sub-task from previous ASVspoof challenges, is predicated on a telephony scenario where synthetic and converted speech is directly injected into a communication system without any acoustic propagation.

The ASVspoof 5 Challenge introduces significant changes in source data, attack types, and evaluation metrics.
The source data, extracted from the Multilingual Librispeech English partition \cite{Pratap2020}, includes a vastly greater number of speakers than previous ASVspoof databases. 
Notably, the spoofing attacks in the \textit{training}, \textit{development}, and \textit{evaluation} sets are entirely disjoint.
As shown in Table \ref{table1}, the \textit{training} dataset is used to adjust model parameters, while the \textit{development} dataset is used to tune and evaluate performance. 
The \textit{progress} set initially assesses the detection model's performance, allowing participants up to four submissions per day via the Codalab platform.
The \textit{evaluation} set tests its generalizability, with only one submission allowed per team.
%
New evaluation metrics, minDCF \cite{Sadjadi2020} for Track 1 and agnostic DCF \cite{Shim2024} for Track 2, have been introduced to better assess anti-spoofing systems.

This paper presents the SZU-AFS anti-spoofing system for Track 1 under open conditions. 
Its design diagram is illustrated in Figure \ref{Fig1}, where model IDs are labeled from A to D, with numbers indicating their respective versions.
This system has four stages: baseline model selection, exploration of effective data augmentation (DA) methods for fine-tuning, application of a co-enhancement strategy utilizing gradient norm aware minimization (GAM) for secondary fine-tuning, and fusion of logits scores from the two top-performing models.
Specifically, comparative experimental analysis was conducted first by combining three pre-trained models with three distinct classifiers to select an appropriate baseline model. 
We have selected the pre-trained Wav2Vec2 model \cite{Baevski2020} as the feature extractor, coupled with the AASIST classifier \cite{Jung2022}, to serve as the baseline model (A9).
Secondly, we have proposed three DA policies to explore the effectiveness of various DA methods: single-DA, random-DA, and cascade-DA. 
The best-performing model is the one fine-tuned by augmented data generated by sequentially applying room impulse response (RIR) noise and time masking (TimeMask) method, resulting in an augmented model (B5).
Next, we employed a GAM-based co-enhancement strategy to consider data and optimizer simultaneously to enhance model generalizability.
With this strategy, the B5 model has been fine-tuned by combining various DA methods with the GAM method, resulting in the C11 and C12 models as the two best-performing fine-tuned models.
Finally, we have fused the predicted logits scores from the C11 and C12 models using an average score-level fusion method to generate final evaluation scores, constituting system D4.

This paper is organized as follows:
Section \ref{Title2} elaborates the core modules of the SZU-AFS system, including the baseline model, the three DA policies, the GAM-based co-enhancement strategy, and the score-level fusion.
Implementation details regarding the dataset information and model hyperparameters are provided in Section \ref{Title3}.
Section \ref{Title4} provides experimental results and analysis. 
Conclusions are drawn in Section \ref{Title5}.

\section{Methodology}\label{Title2}
The SZU-AFS anti-spoofing system consists of four stages detailed in separate subsections: baseline model, data augmentation, gradient norm aware minimization (GAM)-based co-enhancement strategy, and score-level fusion.
Note that we trained ten different baseline models, six augmented models, and eight models using various DA and GAM methods. 
Model IDs are labeled A to D, and numbers indicate versions.


\subsection{Baseline Model}\label{Title2_1}
\subsubsection{Front-end feature extractor}
Given the urgent need to improve the generalizability of spoofing detection systems, speech self-supervised models have gained increasing attention.
Prior research shows that using speech self-supervised models as the front-end feature extractors and the back-end classifier, can substantially improve the generalization of spoofing detection models \cite{MartinDonas2022,Lee2022,Zhong2024,Yang2024}.

We have used the self-supervised WavLM-Base\footnote{https://github.com/microsoft/unilm/blob/master/wavlm} \cite{Chen2022}, HuBERT-Base\footnote{https://github.com/facebookresearch/fairseq/tree/main/examples/hubert} \cite{Hsu2021}, and Wav2Vec2-Large\footnote{https://github.com/facebookresearch/fairseq/tree/main/examples/wav2vec} \cite{Baevski2020} as front-end feature extractors instead of conventional handcrafted acoustic features, such as linear frequency cepstral coefficients and mel-spectrograms.
The self-supervised learning models extract speech representations or embeddings from the raw waveform.

\subsubsection{Back-end classifier}
The back-end classifiers of the latest spoofing detection systems mainly adopt deep learning methods \cite{Wang2023,Rosello2023,Tak2022}, significantly outperforming traditional classifiers such as support vector machine and Gaussian mixture model \cite{Yi2023, Xu2024}.
We have tried three representative classifiers combined with front-end pre-trained models, detailed as follows:
\begin{itemize}
	\item \textbf{Fully connected (FC) classifier} \cite{Wang2023}: This classifier combines a global average pooling layer, followed by a neural network with three fully connected layers employing LeakyReLU activation functions. It ends with a linear output layer for binary classification.
	\item\textbf{Conformer classifier} \cite{Rosello2023}: This classifier combines a convolutional neural network and a Transformer network for spoofing detection. It comprises four blocks, each with four attention heads and a kernel size of 31, totaling 2.4 million parameters.
	\item \textbf{AASIST classifier} \cite{Tak2022}: This classifier combines a RawNet2-based encoder \cite{Tak2021a} and a graph network module. Specifically, it removes the Sinc convolutional layer-based front-end from the RawNet2-based encoder.  
\end{itemize}

\subsubsection{Model selection}
As shown in Table \ref{table2}, we have evaluated the detection performance of the A1-A10 models using the development set of ASVspoof 5. 
The A1-A9 models are combinations of three pre-trained models with three different classifiers, while the A10 model combines the Wav2Vec2 pre-trained model with all classifiers, generating predictive scores by processing concatenated features through a linear layer.
According to experimental results, we have selected the A9 model as the baseline by utilizing a Wav2Vec2-based front-end feature extractor paired with an AASIST-based back-end classifier.

\begin{figure}[t]
	\centering
	\includegraphics[scale=0.16]{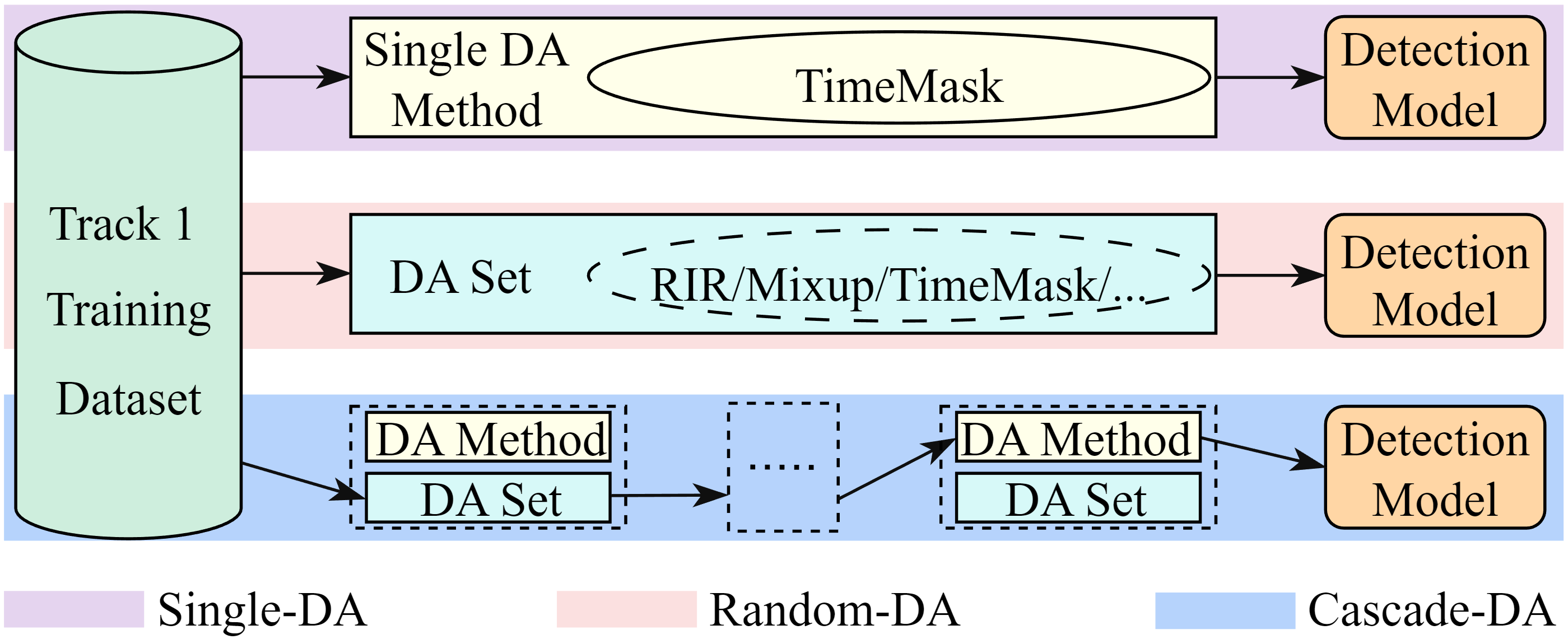}
	\caption{Illustration of the three different DA policies. To enhance the generalization abilities of the A9 model, we experiment with three distinct DA policies, including single-DA, random-DA, and cascade-DA.}
	\vspace{-15pt} 
	\label{spprod}
\end{figure}

\subsection{Primary Fine-tuning with Data Augmentation}\label{Title2_2}
To enhance the generalization performance, we have conducted experiments with three DA policies: single-DA, random-DA, and cascade-DA, to fine-tune the baseline model.
The three DA policies, as depicted in Figure \ref{spprod}, are detailed as follows.

\subsubsection{Single-DA policy}
A specific DA method is applied to all original training data for the single-DA policy. 
The details of the DA methods employed are described below:
\begin{itemize}
	\item \textbf{RIR} \footnote{https://www.openslr.org/28/}: The room impulse response (RIR) captures the acoustic characteristics of a room or an environment. A noise clip is randomly selected from the RIR database and superimposed onto the original training speech, with the intensity randomly varying between 20\% and 80\%.
	\item \textbf{RawBoost} \cite{Tak2022a}: RawBoost incorporates 3 distinct types of noise: linear and non-linear convolutive (LnL) noise, stationary signal-independent additive (SSI) noise, and impulsive signal-dependent additive (ISD) noise.
	\item \textbf{Signal companding}: The a-law and $\mu$-law are signal companding methods developed to enable the transmission of signals with a large dynamic range through systems with limited dynamic range capabilities. During the enhancement of the input speech, either a-law or $\mu$-law is randomly selected.
	\item \textbf{TimeMask}: For the input speech, consecutive time steps from $t_0$ to $t_0 + t$ are set to zero. The duration $t$ is uniformly selected from 0 to $T$, and the starting point $t_0$ is randomly chosen from the interval $[0, \tau - t)$. Here, $\tau$ represents the total number of time steps, and $T$ varies randomly between 20\% and 50\% of $\tau$.
	\item \textbf{Mixup} \cite{Zhang2018}: Mixup regularization involves training the model using a set of mixed speech utterances and labels, rather than the original training data, with the interpolation parameter sampled from a symmetric $Beta(\sigma, \sigma)$ distribution, where $\sigma=1.0$.
	\item \textbf{Amplitude}: Amplitude enhancement involves selecting two speech utterances from the same speaker and label, mixing their amplitude spectra with a certain probability, and then applying inverse Fourier transformation with the corresponding phase spectra to obtain the enhanced utterances.
\end{itemize}

\subsubsection{Random-DA policy}
Unlike a single-DA strategy, the random-DA policy involves randomly selecting an augmentation method from a DA set for each utterance of the original training data.
More specifically, we used three DA sets:
\begin{itemize}
	\item \textbf{Noise set}: This set contains 3 noise-based DA methods from the audiomentations\footnote{https://github.com/iver56/audiomentations} library, with corresponding modules named AddColorNoise, AddGaussianNoise, and AddGaussianSNR.
	\item \textbf{Filter set}: This set contains 7 filter-based DA methods from the audiomentations library, with corresponding modules named BandPassFilter, BandStopFilter, HighPassFilter, HighShelfFilter, LowPassFilter, LowShelfFilter, and PeakingFilter.
	\item \textbf{Mix set}: This set contains 13 DA methods of mixed types from the audiomentations library, with corresponding modules named AddGaussianNoise, AirAbsorption, Aliasing, BandPassFilter, Shift, PitchShift, HighPassFilter, LowPassFilter, PolarityInversion, PeakingFilter, TimeStretch, TimeMask, and TanhDistortion.
\end{itemize}

\subsubsection{Cascade-DA policy}
The cascade-DA policy encourages selecting two or more DA methods in a sequential cascade manner to enhance the original training data progressively.
Three types of cascade-DA methods are given below:
\begin{itemize}
	\item \textbf{RIR-TimeMask}: RIR-TimeMask consists of a two-level cascade of DA methods, sequentially adding RIR noise and TimeMask method to the original training data.
	\item \textbf{LnL-ISD}: LnL-ISD consists of a two-level cascade of DA methods, sequentially adding LnL and ISD noise to the original training data. Both LnL noise and ISD noise are derived from the RawBoost method.
	\item \textbf{Noise-Filter}: Noise-Filter consists of a two-level cascade of DA sets, sequentially applying one method randomly selected from the noise set and another from the filter set to enhance the original training data.
\end{itemize}

Note that the RIRTimeMask method is used in the primary fine-tuning stage, while LnL-ISD, Noise-Filter, and combinations of cascade-DA methods are used in the secondary fine-tuning stage. 

\subsubsection{Model selection}
The A9 model is fine-tuned using only the on-the-fly augmented data.
We have evaluated the detection performance of six models (B1-B6, which are shown in Table \ref{table3}) with different DA methods, using the \textit{progress} set of ASVspoof 5.
Specifically, we have fine-tuned the A9 model using distinct DA methods, including Amplitude, a-law or $\mu$-law, Mix, and RIR-TimeMask.
The B4-B6 models share the same augmentation methods but vary in the number of input speech samples used for training: 64,600, 96,000, and 128,000, respectively.
Following experimental analysis, the B5 model has been chosen for further investigation.

\subsection{Secondary Fine-tuning with GAM-based Co-enhancement Strategy}\label{Title2_3}
Unlike DA methods, which focus on increasing the diversity of training data, the GAM method is an optimization approach for enhancing model generalization.
To alleviate this issue, the fine-tuning process has been divided into two stages: a primary stage without GAM, as described in the previous subsection, and a secondary stage with DA and GAM co-enhancement, as illustrated in this subsection. 

\vspace{-5pt} 
\subsubsection{Gradient norm aware minimization}
Sharpness-aware minimization (SAM) \cite{Foret2021} and its variants \cite{Kwon2021} are representative training algorithms to seek flat minima for better generalization.
Shim et al. \cite{Shim2023} employed SAM and its variants in spoofing detection, improving model generalization.
Inspired by this, we exploit a recently proposed optimization method, gradient norm aware minimization (GAM) \cite{Zhang2023}.

GAM seeks flat minima with uniformly small curvature across all directions in the parameter space.
Specifically, it improves the generalization of models trained with the Adam optimizer by optimizing first-order flatness, which controls the maximum gradient norm in the neighborhood of minima.

Let $\theta \in \Theta \subseteq \mathbb{R} ^d$ denote the parameters of the B5 model. 
The Adam optimizer is then described as follows:
\begin{equation}
	\theta_{t+1}=\theta_{t}-\eta g_t,
	\label{eq3}
\end{equation}
where $t$ is the time step, $\eta$ is the learning rate, and $g_t$ is the loss gradient.
For the first-order flatness $R_{\rho}^{1}(\theta)$, it could be computed by:
\begin{equation}
	 R_{\rho}^{1}(\theta) \triangleq \rho \cdot \max_{\theta' \in B(\theta, \rho)} \|\nabla \hat{L}(\theta')\|, 
	\label{eq5}
\end{equation}
where $\hat{L}(\theta)= {\textstyle \sum_{i=1}^{n}\ell(\theta,x_i,y_i ) }$ denotes the empirical loss function, $x_i$ and $y_i$ denote the $i$-th speech sample and its corresponding label, respectively.  $\rho > 0$ is the perturbation radius that controls the magnitude of the neighborhood, and $B(\theta, \rho)$ denotes the open ball of radius  $\rho$ centered at the parameter $\theta$ in the Euclidean space.
For detailed derivation, see Appendix A of \cite{Zhang2023}.
The key to optimizing generalization error with GAM is controlling the loss function $\hat{L}(\theta)$ and first-order flatness $R_{\rho}^{1}(\theta)$. 
The pseudocode of the whole optimization procedure is shown in Algorithm \ref{alg1}.
\begin{algorithm}[!ht]
	\renewcommand{\algorithmicrequire}{\textbf{Input:}}
	\caption{Gradient norm Aware Minimization (GAM)}
	\label{alg1}
	\begin{algorithmic}[1] 
		\REQUIRE  Batch size $b$, learning rate $\eta_t$, perturbation radius $\rho_t$, trade-off coefficient $\alpha$, small constant $\xi$ 
		\STATE $t \leftarrow 0$, $\theta_0 \leftarrow \text{initial parameters}$
		\WHILE {$\theta_t$ not converged}
		\STATE Sample $W_t$ from the training data with $b$ instances
		\STATE Calculate the empirical loss gradient $\nabla \hat{L}(\theta_t)$: \\ $h^{\text{loss}}_t \leftarrow \nabla \hat{L}(\theta_{t})$ 
		\STATE Calculate the perturbed gradient using the loss gradient \( \nabla \hat{L}_{W_t}(\theta_t) \) of the sample $W_t$ and the Hessian matrix \( \nabla^2 \hat{L}_{W_t}(\theta_t) \):  \\ $f_t \leftarrow \nabla^2 \hat{L}_{W_t}(\theta_t) \cdot \frac{\nabla \hat{L}_{W_t}(\theta_t)}{\|\nabla \hat{L}_{W_t}(\theta_t)\|+\xi}$
		\STATE Calculate the adversarial parameters adjusted via the perturbed gradient: \\ $\theta^{\text{adv}}_t \leftarrow \theta_t + \rho_t \cdot \frac{f_t}{\|f_t\|+\xi}$
		\STATE  Calculate the norm gradient $\nabla R_{{\rho}_t}^{(1)} (\theta_t)$:  \\ $h^{\text{norm}}_t \leftarrow \rho_t \cdot \nabla^2 \hat{L}_{W_t}(\theta^{\text{adv}}_t) \cdot \frac{\nabla \hat{L}_{W_t}(\theta^{\text{adv}}_t)}{\|\nabla \hat{L}_{W_t}(\theta^{\text{adv}}_t)\|+\xi}$
		\STATE $\theta_{t+1} \leftarrow \theta_t - \eta_{t} (h^{\text{loss}}_t + \alpha h^{\text{norm}}_t)$
		\STATE $t \leftarrow t + 1$
		\ENDWHILE
		\STATE \textbf{return} $\theta_{t}$.
		\end{algorithmic}
\end{algorithm}

\vspace{-10pt} 
\subsubsection{Co-enhancement strategy}
The GAM-based co-enhancement strategy involves data augmentation of the input speech and combines the GAM method with the Adam optimizer to further fine-tune the DA-augmented baseline model (B5).
Unlike the primary fine-tuning with DA methods, this strategy has explored more efficient two-level and three-level DA methods, combined with RIR or TimeMask, to process the original training data.
Specifically, we have combined eight different DA methods with GAM: C1 (RIR), C2 (a-law or $\mu$-law), C3 and C4 (Mix), C5 and C6 (LnL-ISD), C7 (RIR + Mix), C8 and C9 (RIR-TimeMask), C10 and C11 (RIR-TimeMask + Mixup), and C12 (RIR + Noise-Filter).
As shown in Table \ref{table4}, we have evaluated the detection performance of models C1-C12 using the \textit{progress} set of ASVspoof 5.
Experimental analysis indicates that the C11 and C12 models are the two best-performing models in terms of minDCF.

\subsection{Score-level Fusion}
The individual model scores have been directly output as logits from the linear layer without applying min-max normalization. 
Building on this, we have utilized an average score-level fusion method, where the predicted scores from each model have been summed and averaged to determine the final prediction score.

As shown in Table \ref{table5}, we have evaluated the detection performance of fused models D1-D4 on either the \textit{progress} or \textit{evaluation} sets of ASVspoof 5.
Specifically, we have tested four fused models: D1 (B4 + B5), D2 (B1 to B6), D3 (C8 + C9), and D4 (C11 + C12).
Among these models, we have selected the best-performing fused system, D4, for submission to the evaluation phase.

\begin{table}[t]
	\caption{\label{table1} {\it Summary of ASVspoof 5 Track 1 database. “Spr.” denotes the number of speakers, while “Train.”, “Dev.”, “Prog.”, “Eval.” refer to the \textit{training}, \textit{development}, \textit{progress}, and \textit{evaluation} sets, respectively.	}}
	\vspace{2mm}
	\centering
	\centerline{
		\begin{tabular}{|c|c|c|c|c|c|}
			\hline
			\multirow{2}{*}{Sets}  & \multirow{2}{*}{Spr.} & \multirow{2}{*}{\begin{tabular}[c]{@{}c@{}}Attack \\ Types \end{tabular} }& \multicolumn{3}{c|}{Utterances} \\ \cline{4-6} 
			&             &      & Bona fide   & Spoofed    & Total    \\
			\hline \hline
			Train.        & 400      & A1-A8    & 18,797      & 163,560  & 182,357  \\
			Dev.       & 785     & A9-A16       & 31,334      & 109,616  & 140,950  \\
			Prog.        & —      & —          & —           & —        & 40,765   \\
			Eval.      & 737     & A17-A32      & 395,924           & 138,688      & 680,774  \\ \hline
	\end{tabular}}
\end{table}

\section{Experimental Setup}\label{Title3}
\subsection{Datasets and Metrics}
\subsubsection{Datasets}
This paper focuses on the Track 1 stand-alone speech deepfake detection task of ASVspoof 5, with a summary of the Track 1 database provided in Table \ref{table1}. 
The dataset contains 1,044,846 utterances, each encoded as a 16 kHz, 16-bit FLAC file. 
The \textit{training} and \textit{development} sets each contain spoofed speech generated by 8 different text-to-speech (TTS) or voice conversion (VC) methods. 
In contrast, the \textit{evaluation} set includes spoofed speech from 16 diverse attack methods, including TTS, VC, and, for the first time, adversarial attacks.
The \textit{evaluation} set contains more than twice the number of samples as the combined \textit{training} and \textit{development} sets, making detection significantly more challenging.
Notably, the \textit{progress} set is a subset of the \textit{evaluation} set.

%


\subsubsection{Metrics}
Different from previous ASVspoof challenges, ASVspoof 5 Challenge uses the minDCF as the primary metric for the comparison of spoofing countermeasures, with the cost of log-likelihood ratio ($C_{llr}$) \cite{Bruemmer2006} and the equal error rate (EER) as a secondary metrics.
Accuracy (ACC) was introduced to evaluate the detection model's performance on the development set. 
In contrast, EER provides a more suitable measure of performance when the data is limited or imbalanced.
Thus, EER is better suited than ACC for evaluating spoof detection models.

The normalised detection cost function (DCF) is: 
\vspace{-5pt} 
\begin{equation}
	\text{DCF}(\tau_{cm}) = \beta \cdot P_{\text{miss}}^{\text{cm}}(\tau_{cm}) + P_{\text{fa}}^{\text{cm}}(\tau_{cm}), 
	\beta = \frac{C_{\text{miss}}}{C_{\text{fa}}} \cdot \frac{1 - \pi_{\text{spf}}}{\pi_{\text{spf}}},
	\label{eq1}
\end{equation}
where $\tau_{cm}$ is the detection threshold, $\pi_{\text{spf}}$ is asserted prior probability of spoofing attack, and $C_{\text{miss}}$ and $C_{\text{fa}}$ are the costs of a miss and a false alarm, respectively.
The following parameters were used for the ASVspoof 5 challenge evaluation: $C_{\text{miss}} = 1$, $C_{\text{fa}} = 10$, $\pi_{\text{spf}} = 0.05$, and $\beta \approx 1.90$.
The normalized DCF in (\ref{eq1}) is used to compute the minimum DCF, defined as $\text{minDCF}= \min_{\tau_{cm}}  \text{DCF}(\tau_{cm})$.

\begin{table}[t]
	\caption{\label{table2} {\it Performance in Accuracy (\%) and EER (\%) of different baseline models on the Track 1 development set. The highlighted model was selected for further fine-tuning to enhance its generalizability.}}
	\vspace{2mm}
	\centering
	\centerline{
		\begin{tabular}{|c|c|c|c|c|}
			\hline
			\begin{tabular}[c]{@{}c@{}}Model \\ ID \end{tabular}  & \begin{tabular}[c]{@{}c@{}}Feature \\ Extractor \end{tabular}   & \begin{tabular}[c]{@{}c@{}}Back-end \\ Classifier \end{tabular} & \begin{tabular}[c]{@{}c@{}}Accuracy \\ (\%) \end{tabular}  & \begin{tabular}[c]{@{}c@{}}EER \\ (\%) \end{tabular}   \\ \hline\hline
			A1        & \multirow{3}{*}{WavLM}       & FC                  & 54.56    & 40.00    \\
			A2        &                              & Conformer           & 67.80     & 43.50  \\
			A3        &                              & AASIST              & 77.25    & 42.70  \\ \hline
			A4        & \multirow{3}{*}{HuBERT}      & FC                  & 73.12    & 19.43 \\
			A5        &                              & Conformer           & 78.31    & 9.56  \\
			A6        &                              & AASIST              & 81.58    & 7.81  \\ \hline
			A7        & \multirow{5}{*}{Wav2Vec2} & FC                  & \textbf{91.49}    & 2.17  \\
			A8        &                              & Conformer           & 81.81    & 6.50   \\
			\cellcolor{pink}A9  &        & \cellcolor{pink}AASIST    & \cellcolor{pink}87.64    & \cellcolor{pink}\textbf{1.55}  \\
			A10 &                              & \begin{tabular}[c]{@{}c@{}}FC + AASIST \\  + Conformer \end{tabular}  & 88.56   & 2.04   \\
			\hline
	\end{tabular}}
\end{table}

\begin{table*}[t]
	\caption{\label{table3} {\it Effect of A9 model with various DA methods on Track 1 \textit{progress} phase. The highlighted model was selected for further fine-tuning to enhance its generalizability.}}
	\vspace{2mm}
	\centering
	\centerline{
		\begin{tabular}{|c|c|c|c|c|c|c|c|c|c|}
			\hline
			\begin{tabular}[c]{@{}c@{}}Model \\ ID \end{tabular}   & \begin{tabular}[c]{@{}c@{}}DA \\ Policy\end{tabular} & DA Method & Optimizer & \begin{tabular}[c]{@{}c@{}} Sample Points \\ of \textit{Training} \end{tabular} & \begin{tabular}[c]{@{}c@{}}Sample Points \\ of \textit{Progress} \end{tabular} & minDCF & actDCF & $C_{llr}$   & EER \\ 
			\hline \hline
			B1 & Single &Amplitude     &    \multirow{6}{*}{Adam}    & 64,600    & 64,600  & 0.137 & 0.322 & 0.466 & 6.76 \\
			B2 & Random & a-law or $\mu$-law  &      & 64,600    & 64,600  & 0.063 & \textbf{0.108}  & 0.287 & 2.32 \\
			B3 & Random & Mix     &       & 64,600    & 64,600  & 0.139 & 0.454  & 0.553 & 6.21 \\
			B4 & Cascade & RIR-TimeMask &       & 64,600    & 64,600  & 0.057 & 0.420 & 1.508 & 2.05 \\
			\cellcolor{pink}B5 & \cellcolor{pink}Cascade & \cellcolor{pink}RIR-TimeMask &       & \cellcolor{pink}96,000   & \cellcolor{pink}96,000  & \cellcolor{pink}\textbf{0.043} & \cellcolor{pink}0.116 & \cellcolor{pink}\textbf{0.235} &\cellcolor{pink} \textbf{1.50}  \\
			B6 & Cascade & RIR-TimeMask &       & 128,000    & 128,000  & 0.067 & 0.143 & 0.302 & 2.46 \\
			\hline
	\end{tabular}}
\end{table*}

\begin{table*}[t]
	\caption{\label{table4} {\it Effect of the B5 model under GAM-based co-enhancement strategy on Track 1 \textit{progress} phase. }}
	\vspace{2mm}
	\centering
	\centerline{
		\begin{tabular}{|c|c|c|c|c|c|c|c|c|c|}
			\hline
				\begin{tabular}[c]{@{}c@{}}Model \\ ID \end{tabular}  & \begin{tabular}[c]{@{}c@{}}DA \\ Policy\end{tabular} & DA Method & Optimizer & \begin{tabular}[c]{@{}c@{}}Sample Points \\  of \textit{Training} \end{tabular}  & \begin{tabular}[c]{@{}c@{}}Sample Points \\  of \textit{Progress} \end{tabular} & minDCF & actDCF & $C_{llr}$   & EER\\  \hline\hline
			C1 & Single & RIR       &    \multirow{12}{*}{\begin{tabular}[c]{@{}c@{}}Adam \\ + \\ GAM\end{tabular}}        &  \multirow{12}{*}{64,600}                    & 96,000  & 0.058 & \textbf{0.062} & 0.111 & 2.07 \\ \cline{1-3}\cline{6-10}
			C2 & Random & a-law or $\mu$-law  & &   & 96,000  & 0.046 & 0.067 & 0.204  & 1.63 \\ \cline{1-3}\cline{6-10}
			C3 & \multirow{2}{*}{Random} & \multirow{2}{*}{Mix}      &  &       & 64,600  & 0.064 & 0.322 & 0.661 & 2.26 \\ 
			C4     &                                  &     &        &             & 96,000  & 0.050 & 0.194 & 0.365 & 1.79 \\ \cline{1-3}\cline{6-10}
			C5 & \multirow{2}{*}{Cascade} & \multirow{2}{*}{LnL-ISD}   &     &      & 64,600  & 0.057 & 0.230 & 0.367 & 2.06 \\
			C6   &              &         &       &               & 96,000  & 0.048 & 0.155 & 0.257 & 1.71 \\ \cline{1-3}\cline{6-10}
			C7 & Cascade & RIR + Mix  &    &  & 96,000  & 0.046  & 0.149 & 0.221 & 1.63 \\ \cline{1-3}\cline{6-10}
			C8 & \multirow{2}{*}{Cascade} & \multirow{2}{*}{RIR-TimeMask} &    & & 64,600  & 0.051 & 0.189 & 0.314 & 1.84 \\
			C9    &                           &      &         &               & 96,000  & 0.041 & 0.190 & 0.276 & 1.48 \\ \cline{1-3}\cline{6-10}
			C10 & \multirow{2}{*}{Cascade} & \multirow{2}{*}{RIR-TimeMask + Mixup} &    &    & 64,600  & 0.050   & 0.922 & 1.688 & 1.77 \\ 
			C11   &                &      &       &             & 96,000  & 0.038 & 0.840 & 1.334 & 1.39 \\ \cline{1-3}\cline{6-10}
			C12 &Cascade & RIR + Noise-Filter & &  & 96,000  & \textbf{0.035} & 0.087 & \textbf{0.108} & \textbf{1.30}  \\ 
			\hline
	\end{tabular} }
\end{table*}

\subsection{Training Details}
In our experiments, the following parameters were kept consistent. 
We used the Adam optimizer $(\beta_1=0.9,\ \beta_2=0.999,\ \epsilon=10^{-8})$ \cite{Kingma2015}, with an initial learning rate of $5 \times 10^{-6}$, controlled by a cosine annealing scheduler with a minimum learning rate of $1 \times 10^{-8}$, and a maximum of 100 training epochs. 
The training was conducted using conventional cross-entropy loss, with early stopping applied if the development set loss did not improve within ten epochs.
All experiments were executed on two NVIDIA A100 GPUs. 
The training epochs for the models used to obtain the D4 system were as follows: 12 epochs (A9),  4 epochs (B5),  2 epochs (C11), and 5 epochs (C12).
The training time required for the combined DA and GAM method is approximately three times that of the regular DA method alone.
The results table highlights the best-performing values in bold for each column.

\section{Results and Analysis}\label{Title4}
\subsection{Comparison Analysis of Different Baseline Models}
We integrated three pre-trained models with three classifiers to assess the necessity of different baseline models, testing their performance on the Track 1 \textit{development} set. 
The training and development data were truncated or padded to 64,600 sample points to accommodate GPU memory constraints.

The accuracy and EER for different baseline models are reported in Table \ref{table2}. 
As indicated in Table \ref{table2}, we observe:
\begin{itemize}
	\item Among the different feature extractors, Wav2Vec2-based detection models (e.g., A7, A8, A9, and A10) achieved higher accuracy and lower EER, which indicates their better effectiveness than WavLM-based and HuBERT-based detection models. 
	This result is due to the fact that both the WavLM and HuBERT pre-trained models use the base version, which contains fewer than one-third of the learnable parameters in the Wav2Vec2-Large model.
	\item 
	Among the different classifiers, AASIST proved to be more competitive than others (e.g., FC and Conformer) when paired with various feature extractors, further confirming its excellent performance in speech spoofing detection.
	Furthermore, leveraging the Wav2Vec2 feature extractor, we concatenated the final predictive outputs from three classifiers for classification.
	However, the A10 model's results fell between the individual classifiers' results, providing no improvement.
	Thus, using only the best classifier is sufficient.
	\item While the A9 model's accuracy is slightly lower at 87\% compared to the A7 and A10 models, it achieves the lowest EER at 1.55\%.
	Owing to its outstanding EER performance, the A9 model has been selected for further experimental exploration.
\end{itemize}

\begin{table}[th]
	\caption{\label{table5} {\it The performance of different fused systems was evaluated on the ASVspoof 5 Track 1 database. “Prog.” and “Eval.” refer to the \textit{progress} and \textit{evaluation} sets, respectively.}}
	\vspace{2mm}
	\centering
	\centerline{
		\begin{tabular}{|c|c|c|c|c|c|c|}
			\hline 
			Phase & ID    & System           & minDCF & actDCF & $C_{llr}$  & EER\\ \hline\hline
			\multirow{4}{*}{Prog.} &D1 &  B4 + B5   & 0.039 & 0.307  & 0.635 & 1.33 \\
			&D2&  B1 $\sim$ B6   & 0.037  & \textbf{0.167}  & \textbf{0.305} & 1.31 \\
			&D3&  C8 + C9   & 0.040 & 0.633  & 0.456 & 1.41 \\
			&D4& C11 + C12   & \textbf{0.027} & 0.269  & 0.366 & \textbf{0.99} \\ \hline
			Eval.  &D4& C11 + C12   & 0.115 & 0.573  & 0.956 & 4.04 \\
			\hline
	\end{tabular}}
\end{table}

\vspace{-10pt} 
\subsection{Comparison Analysis of Different DA}
Table 3 shows the effectiveness of various DA methods during the progress phase of Track 1.
Compared with the B1 and B3 models, the B2 and B4 models achieved lower minDCF and EER.
The A9 model presents superior detection performance when fine-tuned with signal compression (a-law or $\mu$-law), RIR noise, and TimeMask.
However, the B4 model exhibited a higher $C_{llr}$.
The B5 model outperforms all other models in terms of minDCF, $C_{llr}$, and EER at 0.043, 0.235, and 1.5\%, respectively.
Owing to its outstanding performance, the B5 model has been selected as the augmented model for further experimental exploration.

Although current experimental results do not conclusively determine which of the three different DA policies is most effective.
We recommend prioritizing experimental exploration under random-DA and cascade-DA policies for speech spoofing detection tasks.

\subsection{Effect of GAM-based Co-enhancement Strategy}
With the B5 model's good results, we also investigate whether the spoofing detection performance can be further improved by using the GAM method.
Table \ref{table4} shows the performance of various DA methods and GAM method on the Track 1 progress phase.

For the effect of data augmentation, the C1 and C2 models did not significantly improve minDCF and EER over the B5 model in the progress phase.
Specifically, the B5 model using RIR-TimeMask (C8 and C9 models) and its combination with the Mixup (C10 and C11 models) outperformed the C2 and C3 models across most metrics, indicating that more complex augmentation can be learning more robust features. 
In addition, the comparison among models from C8 to C11 shows that the Mixup method significantly improves both minDCF and EER, which suggests that it contributes to the improvement's generalizability.
The GAM method, particularly in B5 and C9 models, improved minDCF and EER, effectively enhancing model generalizability.
The experimental results demonstrate the importance of selecting appropriate data augmentation and optimization techniques to enhance spoofing detection performance.

\subsection{Comparison Analysis of Different Fused Systems}
Table \ref{Title5} shows the performance of the four fused systems on either the \textit{progress} or \textit{evaluation} sets of ASVspoof 5.
We observed that score-level average fusion enhances model performance compared to individual detection models, particularly in minDCF and EER metrics.
Fusing C11 and C12 models (D4) resulted in optimal progress phase performance, achieving a minDCF of 0.027 and an EER of 0.99\%.
However, the D4 system exhibited a significant performance discrepancy between the progress and evaluation phases, highlighting the challenging nature of the \textit{evaluation} set.

\subsection{Impact of Different Sample Points}
Table \ref{Title3} also shows a comparison in terms of sample points for the model training. 
Using 96,000 sample points of input speech, the B5 model achieved a lower actDCF and $C_{llr}$ than B4 and B6.
The B6 model exhibited poor performance, indicating that increasing the number of training samples does not necessarily enhance the model's detection capabilities.
In fact, inputting more sample points for training may reduce the model's generalization ability.
Optimizing training with an appropriate number of sample points is more beneficial for improving detection performance than simply increasing the amount of training data. 

The results presented in Table \ref{Title4} reveal that when the model was trained using input speech with 64,600 sample points, a significant performance improvement was observed during the inference stage when utilizing input speech with 96,000 sample points.
This phenomenon may be associated with the different utterance duration distribution in the \textit{progress} set. 
More studies are required to verify this relationship further and analyze it.




\section{Conclusion}\label{Title5}
This paper describes the SZU-AFS system for Track 1 of the ASVspoof 5 Challenge under open conditions.
Instead of focusing on various pre-trained feature fusion and complex score fusion methods, we used DA and GAM enhancement strategies to improve spoofing detection generalization.
The final best fused system submitted achieved 0.115 minDCF and 4.04\% EER on the ASVspoof 5 challenge \textit{evaluation} set. 

The experiments produced a few valuable findings. 
First, applying the RIR-TimeMask method for data augmentation has proven more effective.
Building on this, employing a cascade-DA strategy can further improve model performance.
Second, the GAM method significantly improves model generalization when combined with the Adam optimizer on both \textit{progress} and \textit{evaluation} sets despite the lengthy training time required.
Due to time constraints, the model was fine-tuned in two stages. 
Using the GAM method throughout the entire process might have produced better results.

\section{Acknowledgments}
We would like to thank the organizers for hosting the ASVspoof 5 Challenge. 
This work was supported in part by NSFC (Grant U23B2022, U22B2047) and Guangdong Provincial Key Laboratory (Grant 2023B1212060076). 
\bibliographystyle{IEEEbib}
\bibliography{ASVspoof_BibEntries}

\end{document}